\documentclass[prb,showpacs,twocolumn,aps,a4paper]{revtex4}
\usepackage{amsmath}
\usepackage{graphicx}
\usepackage{latexsym}
\usepackage{amsfonts}
\usepackage{amssymb}
\DeclareGraphicsExtensions{.pdf,.gif,.jpg}

\newcommand{\be}{\begin{equation}}
\newcommand{\ee}{\end{equation}}
\newcommand{\beq}{\begin{eqnarray}}
\newcommand{\eeq}{\end{eqnarray}}

\begin{document}

\def\bbe{\mbox{\boldmath $e$}}
\def\bbf{\mbox{\boldmath $f$}}
\def\bg{\mbox{\boldmath $g$}}
\def\bh{\mbox{\boldmath $h$}}
\def\bj{\mbox{\boldmath $j$}}
\def\bq{\mbox{\boldmath $q$}}
\def\bp{\mbox{\boldmath $p$}}
\def\br{\mbox{\boldmath $r$}}

\def\bone{\mbox{\boldmath $1$}}

\def\dr{{\rm d}}

\def\tb{\bar{t}}
\def\zb{\bar{z}}

\def\tgb{\bar{\tau}}

\def\bC{\mbox{\boldmath $C$}}
\def\bG{\mbox{\boldmath $G$}}
\def\bH{\mbox{\boldmath $H$}}
\def\bK{\mbox{\boldmath $K$}}
\def\bM{\mbox{\boldmath $M$}}
\def\bN{\mbox{\boldmath $N$}}
\def\bO{\mbox{\boldmath $O$}}
\def\bQ{\mbox{\boldmath $Q$}}
\def\bR{\mbox{\boldmath $R$}}
\def\bS{\mbox{\boldmath $S$}}
\def\bT{\mbox{\boldmath $T$}}
\def\bU{\mbox{\boldmath $U$}}
\def\bV{\mbox{\boldmath $V$}}
\def\bZ{\mbox{\boldmath $Z$}}

\def\bcalS{\mbox{\boldmath $\mathcal{S}$}}
\def\bcalG{\mbox{\boldmath $\mathcal{G}$}}
\def\bcalE{\mbox{\boldmath $\mathcal{E}$}}

\def\bgG{\mbox{\boldmath $\Gamma$}}
\def\bgL{\mbox{\boldmath $\Lambda$}}
\def\bgS{\mbox{\boldmath $\Sigma$}}

\def\bgr{\mbox{\boldmath $\rho$}}

\def\a{\alpha}
\def\b{\beta}
\def\g{\gamma}
\def\G{\Gamma}
\def\d{\delta}
\def\D{\Delta}
\def\e{\epsilon}
\def\ve{\varepsilon}
\def\z{\zeta}
\def\h{\eta}
\def\th{\theta}
\def\k{\kappa}
\def\l{\lambda}
\def\L{\Lambda}
\def\m{\mu}
\def\n{\nu}
\def\x{\xi}
\def\X{\Xi}
\def\p{\pi}
\def\P{\Pi}
\def\r{\rho}
\def\s{\sigma}
\def\S{\Sigma}
\def\t{\tau}
\def\f{\phi}
\def\vf{\varphi}
\def\F{\Phi}
\def\c{\chi}
\def\w{\omega}
\def\W{\Omega}
\def\Q{\Psi}
\def\q{\psi}

\def\ua{\uparrow}
\def\da{\downarrow}
\def\de{\partial}
\def\inf{\infty}
\def\ra{\rightarrow}
\def\lra{\leftrightarrow}
\def\bra{\langle}
\def\ket{\rangle}
\def\grad{\mbox{\boldmath $\nabla$}}
\def\Tr{{\rm Tr}}
\def\Re{{\rm Re}}
\def\Im{{\rm Im}}

\title{Magnetization Transfer  by a Quantum Ring Device  }


\author{Michele Cini}

 \affiliation{ Dipartimento di Fisica, Universit\`{a}
di Roma Tor Vergata, Via della Ricerca Scientifica 1, I-00133 Rome,
Italy, and Istituto Nazionale di Fisica Nucleare - Laboratori
Nazionali di Frascati, Via E. Fermi 40, 00044 Frascati, Italy.}

\author{Stefano Bellucci}
 \affiliation{  Istituto Nazionale di Fisica Nucleare - Laboratori
Nazionali di Frascati, Via E. Fermi 40, 00044 Frascati, Italy.}
\date{\today}

\begin{abstract}
We show that a tight-binding model device consisting of a laterally connected ring at half filling  in a tangent time-dependent magnetic field can in principle be designed to pump a purely spin current. The process exploits the spin-orbit interaction in the ring. This behavior is understood  analytically and found to be robust with respect to temperature and small deviations from half filling.
\end{abstract}


\maketitle

Can we transfer magnetization between two distant bodies directly through a lead, without moving any charges?  We need a  device that can produce a pure spin current, without any charge current associated to it. In this work we show that in principle Quantum Mechanics allows us to build a device to achieve just that, by a laterally connected ring (Fig. 1a)  with  half-filled band and spin-orbit interaction  and a time-dependent magnetic field in the plane of the ring. Due to the  spin-orbit interaction, opposite  spin  electrons circulate in the ring with opposite chiralities, but then the  in-plane time-dependent magnetic field flips  spins and pumps them out, yielding a  pure spin current in both wires. A current of spin-up electrons hops to the right and an equal current of spin-down electrons hops to the left. Taking advantage of such a mechanism, one might realize a device allowing for the magnetization of systems, in a situation in which no capacitor charging can occur. Such an application of our results, if realized in a practical device, would of course offer a very good solution for the problems connected to Spintronics applications. 

In the past several years there has been in the literature a growing interest in persistent as well as transient
currents in quantum rings threaded by a magnetic flux, with a promising outlook in the quest for new device applications in spintronics, memory devices, optoelectronics, quantum pumping, and quantum information processing \cite{devices1, devices2, devices3, devices4}.

Aharonov-Bohm-type thermopower oscillations of a quantum dot embedded in a ring for the case when the interaction between electrons can be neglected, were investigated in the literature, showing it to be strongly flux- and experimental geometry- dependent.\cite{saro1} Also, the general subject of pumping phenomena in mesoscopic or ballistic conductors has been already addressed by several authors of theoretical papers.\cite{pump1, pump2, pump3, pump4, pump5, pump6}. Although electron-electron interactions are not needed, in order to introduce the notion of pumping and study the corresponding phenomena, nonetheless including such interactions in this problem, would yield the breakdown of the Fermi liquid, thus leading to the formation of the Luttinger liquid.\cite{ll1} Within the pumping context, a distinct place was attributed to the pumping properties of a Luttinger liquid\cite{ll2, ll3, ll4} and, in particular, of a quantum ring laterally connected to open one-dimensional leads described within the Luttinger liquid model. \cite{ll5} In a laterally connected ring the external circuit  is tangential to the ring and, in such a maximally asymmetric situation,  a current in the wires  produces a magnetic moment.

 The latter is not obtained by substituting the quantum current in the semiclassical formula. Indeed, in previous works \cite{ciniperfettostefanucci, ciniperfetto} it was shown that in nanoscopic circuits containing loops, magnetic moments excited by currents are dominated by quantum effects and depend nonlinearly on the exciting bias, quite at variance from classical expectations of a linear behavior.

   The creation of a magnetic dipole by a bias-induced current is a process which can be  reversed,by magnetically exciting the ring in the absence of bias. Hence, ballistic rings asymmetrically connected to wires and excited by a time-dependent inner magnetic flux can produce ballistic currents in the external wires even in the absence of an external bias and thus, by the same token, they can be useful, in order to obtain charge pumping. Several methods were found to work, one method being based on the introduction of integer numbers of fluxons, another method consisting in connecting the ring to a junction. In general, by studying the real-time  quantum evolution of tight-binding models in different geometries, several kinds of crucial experimental tests of these ideas can be envisaged, resulting in potentially useful devices. As a direct consequence of the above mentioned nonlinearity, one can achieve, by employing suitable flux protocols, single-parameter nonadiabatic pumping, where an arbitrary amount of charge can be transferred from one side to the other, a phenomenon which, for a linear system, would be readily ruled out by the Brower theorem\cite{brower}.

After this introduction and background description, we next proceed to state the model Hamiltonian (see Fig.1a) as:
\be H=H_{D}+H_{B}\ee
where $H_{D}$ is the device Hamiltonian and $H_{B}$ the magnetic term.
\be
H_{D}=H_{wires}+H_{ring}+H_{ring-wires}.\ee
the $N$-sided ring is represented by
\be
 H_{ring}=t_{ring}\sum_{\sigma}\exp[i\sigma\alpha]c^{\dagger}_{i+1,\sigma}c_{i,\sigma}+h.c.\ee
 Here, following A. A. Zvyagin\cite{Zvyagin}, we included the spin-orbit interaction as a phase  shift  $\alpha$ for up-spin  and $-\alpha_{SO}$ for down-spin electrons. Both wires are modeled by $H_{wires}=H_{L}+H_{R}=t_{h}\sum_{n,\sigma} c^{\dagger}_{n,\sigma}c_{n+1,\sigma},$ and the ring-wires contacts are modeled in $H_{ring-wires}$ whereby the ring is connected
to the leads via a tunneling Hamiltonian with hopping $t_{lr}$
connecting two nearest-neighbor sites of the ring denoted with
A and B with the ending sites of lead L and R, respectively
( Fig. 1a). Below we assume for the sake of definiteness  that $t_h=t_{ring}=t_{lr}=1$ eV.  At equilibrium the occupation of the system is determined by the spin-independent chemical potential $\mu$, which is assumed to be zero (i.e. we assume half filling).
 The ring is taken in the x-y plane and the spin-polarized current is excited by  a time-dependent  external magnetic field $B(t)$  along the x axis.   The magnetic interaction is
\be H_{B}=V(t) \sum_{i \in ring}(c^{\dagger}_{i,\uparrow}c_{i,\downarrow}+ c^{\dagger}_{i,\downarrow}c_{i,\uparrow}   )\ee where $V=\mu B$
with the Bohr magneton $\mu=5.79375 10^{-5}\frac{eV}{Tesla}.$


Initially the system is in the ground state with $V=0$. In order to describe its evolution we need the retarded Green's function matrix elements on a spin-orbital basis:
\be
g^{r}_{i,j}=\langle i|U_I (t,0)|j\rangle \ee
where $U_I (t,0)$ is the evolution operator in the interaction representation; the number current\cite{cini80} is
\be J_{n,\sigma} (t)=-2 \frac{t_h}{h} Im ( G^{<}_{n,\sigma,n-1,\sigma})\ee
where
\be  G^{<}_{i,j}(t)=\sum_\mu n^{0}_\mu g^{r}_{i,\mu}(t,0)g^{r *}_{j,\mu}(t,0).
\end{equation}
where $\mu$ denotes the ground state spin-orbitals for $B=0$ and $n^{0}_\mu$ is the Fermi function.


The analysis of the time evolution  is enormously simplified and can be carried out with generality for any $V(t)$  since the model is bipartite  (i.e. bonds connect sites of two disjoint sublattices), and  can be mapped on a spin-less model which is also bipartite (Fig.1 b). A Dirac monopole (the star in Fig.1 b) ensures the spin-orbit interaction, imparting opposite chirality to the two sub-clusters. Due to the spin-orbit interaction the parity $P: x \rightarrow -x$ and the reflection $\Sigma$ which sends each sub-cluster to the other fail to commute with $H$, but the product $P\Sigma$ is a symmetry. Considering sites at the same distance from the ring on both leads  and using the correspondence  $\alpha \rightarrow $ left, spin up; $\beta \rightarrow $ left, spin down;$\gamma \rightarrow $ right, spin up, and $\delta \rightarrow $ right, spin down,  at any time the currents  are constrained by   $J_\alpha =-J_\delta$ and $J_\beta = -J_\gamma $ and the charge densities obey $\rho_\alpha=\rho_\delta,\rho_\beta=\rho_\gamma .$
\begin{figure}[]
\includegraphics*[width=.28\textwidth]{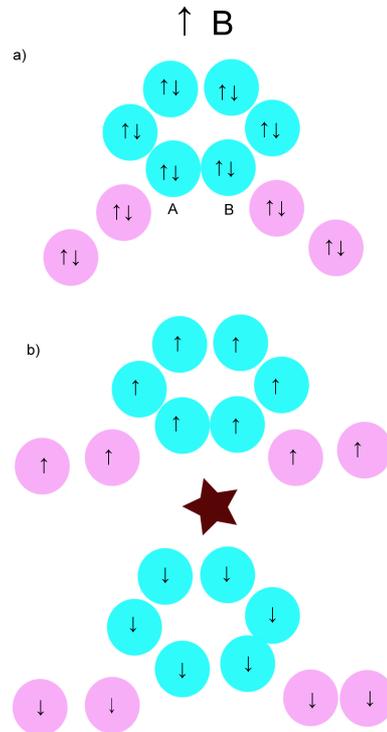}
\caption{Maximally asymmetric connection of the N=6  ring to wires. The circles with  up and down arrows represents the sites; those in the hexagon feel the spin-orbit interaction and the magnetic field. a) Geometry of the device and the magnetic field; all sites are connected horizontally to the first neighbors by spin-diagonal matrix elements, and $B$ flips spins in the hexagon. b) Equivalent cluster for spinless electrons. The star represents the Dirac monopole providing the effective spin-orbit interaction. All sites are connected horizontally to the first neighbors, and those in the ring have also vertical bonds due to the  $V$ magnetic interactions. }
\label{ringgen}
\end{figure}

\begin{figure}[]
\includegraphics*[width=.24\textwidth]{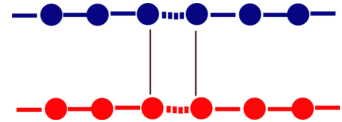}
\caption{The simplified version of our model used to derive Eq. (14). The top (bottom) circles represent the up (down) spin states of a chain; in both horizontal lines the continuous lines stand for identical real hopping matrix elements $t_h$ while the dotted lines represent $t_h e^{i\alpha}$ (top) and $t_h e^{-i\alpha}$ (bottom) connecting, say, sites 0 and 1. The vertical lines stand for $V(t)$ time-dependent hoppings that replace the magnetic interactions in the simplified model. }
\label{ringgen}
\end{figure}

\begin{figure}[]
\includegraphics*[width=.28\textwidth]{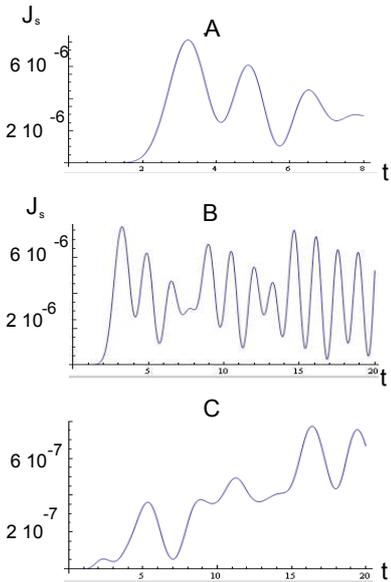}
\caption{Numerical results  of the  model of Eq. (1) with 100 sites in the ring and 100 sites in each lead. The currents are excited by the sudden switching of a field B=100 Tesla and are quadratic with B.
Panel A: spin current $J_{s}$ at half filling and zero temperature. The results hardly change if one takes a temperature of 0.025 $t_{h}$, of the order of room temperature if $t_{h}\sim 1$eV. In both cases the charge current vanishes exactly.
Panel B: on a different time scale, the spin current assuming zero temperature and a Fermi energy 0.01$t_{h}.$ The spin current does not change much. However the charge current $J_{c}$ (Panel C) does not vanish any more, although it is still an order of magnitude smaller than the spin current. }
\label{ringgen}
\end{figure}


The study is simplest in the equivalent lattice of Fig. 1b).  Let us consider first any eigenstate  of the instantaneous H, thought of as stationary. Changing sign to all the amplitudes in a sublattice, we get a solution of the one-electron problem with opposite hopping $t_{h}$, and also a solution of the same Schr\"odinger equation with opposite energy. Hence, if $\epsilon$ is an energy eigenvalue, $-\epsilon$ also is, and opposite energy eigenstates must have the same probability on site.   Coming to the many-body state, the sum of the probabilities is exactly one half. In other terms, each site of the equivalent lattice  is exactly half filled, and in the original model the two spin states are exactly half filled on every site. This holds for any static $B$ including  the initial state where  $B=0$. \\

In the adiabatic limit the system is in the instantaneous ground state at each time. Then, no charge current is allowed, because  the total occupation of each site in Fig. 1a) is fixed; moreover no spin current is allowed either, since it would alter  the  occupation of the sites in Fig. 1b), which is also bipartite. Therefore the adiabatic evolution of this system is trivial. Since we are interested in the non-adiabatic evolution, the beautiful analysis of Ref.\cite{adiabatic} does not apply here.
\\


To show that during the time evolution  $B(t)$ produces a pure spin current in the half filled  system, we change to a staggered spin-reversed  hole representation with $c^{\dagger}_{i,\sigma}=s b_{i,-\sigma},$ where $s=1$ in one sublattice and $s=-1$ in the other. The transformation of any bond goes as follows:
$t^{\sigma}_{n,m}c^{\dagger}_{n,\sigma}c_{m,\sigma} \rightarrow -t^{\sigma}_{n,m}b_{n,-\sigma}b^{\dagger}_{m,-\sigma}$. Since $t^{\sigma}$ is Hermitean this is the same as   $t^{\sigma *}_{m,n}b^{\dagger}_{m,-\sigma}b_{n,-\sigma}$ and since opposite spins have conjugate hoppings  we may rewrite this as $t^{-\sigma}_{m,n}b^{\dagger}_{m,-\sigma}b_{n,-\sigma}.  $ On the other hand, in the ring the sites coupled by $V$ belong to opposite sublattices and so the transformation gives:

\be H_{B}\rightarrow V(t) \sum_{i \in ring}(b^{\dagger}_{i,\uparrow}b_{i,\downarrow}+ b^{\dagger}_{i,\downarrow}b_{i,\uparrow}   ).\ee

In this way,  at every time $t$ the transformed hole Hamiltonian $\tilde{H}(b,b^{\dagger})$ depends on $b$ operators exactly as the original Hamiltonian
 $H(c,c^{\dagger})$ depends on the  $c$ operators. In both pictures the evolution starts in the ground state at half filling  and evolves in the same way. Therefore,  at any time and   for any site $n$,
 \be \langle b^{\dagger}_{n,\sigma}(t) b_{n,\sigma}(t)\rangle_{b} = \langle c^{\dagger}_{n,\sigma} (t) c_{n,\sigma}(t)\rangle_{c}. \ee Here  the l.h.s. is the average at time $t$ in the $b$ picture while the r.h.s. is averaged  at time $t$ in the $c$ picture.
 Hence, operating the canonical transformation on the l.h.s.,
\be  \langle 1-n_{n,-\sigma}(t)\rangle_{c} = \langle n_{n,\sigma}(t)\rangle_{c}, \ee
which implies that the mean total occupation of each site is conserved. This cannot be true if charge currents exist. Indeed, let us consider  the operators straddling  each bond: since at each time
 \be \langle b^{\dagger}_{n+1,\sigma} b_{n,\sigma}\rangle_{b} = \langle c^{\dagger}_{n+1,\sigma} c_{n,\sigma}\rangle_{b} \ee
we may conclude that
 \be \langle c^{\dagger}_{n,-\sigma} c_{n+1,-\sigma}\rangle_{c} = \langle c^{\dagger}_{n+1,\sigma} c_{n,\sigma}\rangle_{c}. \ee
Hence the current is pure spin current, q.e.d.


An analytic formula for the current is desirable,  but  the perturbation treatment in the small parameter $\frac{V}{t_{h}}$  although elementary, is too involved to be enlightening. In order to achieve  a simple  estimate of the effect, and capture the essential mechanism producing the spin current, we replace the ring by a  renormalised bond,
with hopping $t_{h}\rightarrow \tau \exp(i \beta_{\sigma}(t)$ with  $\tau\sim t_{h},$ which implies   an effective  potential drop across the bond which produces the current. Indeed,  in terms of the Peierls prescription, this implies  a spin-dependent electric field $\overrightarrow{E}_{\sigma} $ such that  $\dot{\beta}_{\sigma}=\frac{2\pi}{h}\int e\overrightarrow{E}_{\sigma} d\overrightarrow{l}$.
   The phase and the  effective potential are spin-dependent and produce the spin current.

  As a preliminary, in order to motivate the renormalised bond idea, let us consider the simpler problem of  spinless electrons in the same device, but with a normal magnetic field producing a flux in the ring. Such a model was studied previously\cite{ciniperfetto}; it was shown that by suitable protocols one can insert an integer number of fluxons  in the ring in such a way that the electronic system in the ring  is not left charged and is not excited, while  charge is pumped in the external circuit. Writing the number current in units of  $\frac{t_{h}}{h}$ it turns out that $\int Jdt$ is of order unity for every fluxon.  In this case, the ring is equivalent to an effective  bond  with hopping $t_{h}\rightarrow t_{h} \exp(i \beta(t)).$ The time dependent vector potential entails the effective  bias across the bond is $e \phi_{eff}=\hbar \dot{\beta}.$   The quantum conductivity of the wire was discussed elsewhere\cite{cini80}; at small $\phi_{eff},$ the number current is $J=-\frac{\phi_{eff}}{\pi \hbar}.$ Integrating over time,
  one finds  that the total charge pumped when a fluxon is swallowed by the ring is $Q=\int{Jdt}={\beta\over\pi}\sim 1$. In other terms, inserting a flux quantum in the ring  we shift an electron in the characteristic hopping time of the system. This simple argument is in good agreement with the numerical results\cite{ciniperfetto}.

In  the case with spin, the above approach leads us to change the equivalent model of Figure 1b) to the simplified model of Figure 2, where the vertical bonds again stand for $V(t)$ and the effective bond bears a spin-orbit induced static phase $ \alpha$  which produces no effect at all for  $V$=0. When $V(t)$ is on, however, the electron wave function in the upper wire can interfere with a time dependent contribution from   the opposite spin sector where the phase shift is opposite. Effectively this works like a time dependent phase drop across the upper bond, and an opposite phase drop across the lower one. In first-order perturbation theory  the amplitude to go from $k_{1+}$ in the upper wire to $k_{2-}$ in the upper wire reads
 \be c_{\alpha}(k_{1}, k_{2},t)=\frac{-i}{\hbar}\int_{0}^{t}d\tau e^{i\omega(k_{2},k_{1})}[1+e^{-2i\alpha+i\D_{12}}] V(\tau)\ee
 where $\omega(k_{2},k_{1})=2(\cos(k_{2})-\cos(k1))$ and $\D_{12}=k_{1}-k_{2}.$
 Since the graph of Fig.2 is also bipartite, the current $J$  is spin-dependent and site-independent, and  $J\psi_{k}={2 t_{h}\over \hbar}\psi_{k}.$ Therefore the mean current on the top wire is obtained by summing over occupied states:
 $ \langle J_{\alpha} \rangle= {2 t_{h}\over \hbar}\sum_{\k_{1}}^{\cos(k_{1})<0}\sum_{\k_{2}}^{\cos(k_{2})>0}\sin(k_{2})| c_{\alpha}(k_{1}, k_{2},t)|^{2}. $ The spin current is $J_{s}(\alpha)=J_{\alpha}-J_{-\alpha}.$ The calculation is completed most simply by taking $V(t)=V\theta(t)\theta(T-t)$ with $V=\mu B$  and then letting $T=\frac{\hbar}{t_{h}}$ (short rectangular  spike)
 with the result that   \be J_{s}(\alpha)\sim \frac{t_{h}}{\hbar}\frac{\sin(\alpha)}{2\pi}(\frac{VT}{\hbar})^2.\ee

The numerical results of Fig.3 were  computed for the full model according to Equation (7) by evolving the quantum state by a time-slicing integration of the Schr\"odinger equation. They strikingly  illustrate the above analytic  findings. For any time dependence of $V(t)$ the charge current vanishes identically at half filling. Here we present the results for the case of a sudden switching of $B$. Our codes calculate number currents taking $t_h =1.$ If this is interpreted to mean that $t_h =1 $eV, which corresponds to the frequency $2.42 *10^{14}$ $s^{-1},$ a current $J=1$ from the code means $2.42 *10^{14}$ electrons per second, which corresponds to a charge current of  $3.87*10^{-5}$ Ampere.

We also tested the validity of the simple approximation of  Eq. (14) compared to the full model. For $B=100$  Tesla and $\alpha=1$ one finds $ J_{s}=5*10^{-6}\frac{t_{h}}{\hbar}$ The numerical response to a narrow delta-like  spike yields  $ J_{s}=6*10^{-6}\frac{t_{h}}{\hbar}$ and the quadratic dependence on $B$ is fully confirmed. So the simple approximation works for the full model.
 Finite temperatures do not change significantly the results up to $K_{B}T \sim 0.025$ eV. This is interesting since up to now, strongly spin-polarized  currents have been created and detected in ultra-cold atomic gases only\cite{sommer}. Instead, the results are sensitive to the filling, but for   concentrations  of the order of 0.51   one  gets a spin current with a  small charge current while the ring gets charged.\\

 The present model neglects electron-electron interactions, but  it is clear physically  that adding to the Hamiltonian a correlation term like $U(\hat{n}_{\uparrow}+\hat{n}_{\downarrow}-1)^2$ would tend to reinforce the charge confining  effects described here; in the Hartree approximation, however, it would change nothing since  its average at half filling  vanishes strictly during the evolution of the system.

In conclusion, we presented the theoretical analytical description of a tight-binding model device consisting of a laterally connected ring at half filling in a tangent time-dependent magnetic field that can in principle be designed to pump a purely spin current. The process exploits the spin-orbit interaction in the ring, without which the effect would not occur. This behavior is found by our calculations to be robust with respect to temperature and small deviations from half filling. Our analytical treatment  revealed unusual  physical properties, with potential applications to spintronics.

A wealth of experimental results have been already obtained since the first small quantum rings were fabricated
by self-assembled growth of InAs on GaA,\cite{exp1, exp2} but maybe the best is yet to come!
\section{Acknowledgements} The authors are grateful to Matteo Colonna for help with a computer code during the early stages of this project.
%


\end{document}